\documentclass[noshowpacs,amsmath,
twocolumn,
superscriptaddress,
8pt,
aps,prb]{revtex4-1}
\usepackage{setspace}
\usepackage{amsmath}
\usepackage{graphicx}
\usepackage[nearskip,margin = 0pt]{subfig}

\usepackage{verbatim}
\usepackage{amsfonts}
\usepackage{amssymb}
\usepackage{epstopdf} 
\usepackage{xcolor}
\DeclareGraphicsExtensions{.pdf,.eps,.png,.jpg,.mps}

\begin{document}

\title{Generation of high-stability solitons at microwave rates on a silicon chip}

\author{Xu Yi$^{\ast}$, Qi-Fan Yang$^{\ast}$, Ki Youl Yang$^{\ast}$, Myoung-Gyun Suh, and Kerry J. Vahala$^{\dagger}$\\
T. J. Watson Laboratory of Applied Physics, California Institute of Technology, Pasadena, California 91125, USA.\\
$^{\ast}$These authors contributed equally to this work.\\
$^{\dagger}$Corresponding author: vahala@caltech.edu}

\maketitle


{\bf Because they coherently link radio/microwave-rate electrical signals with optical-rate signals derived from lasers and atomic transitions, frequency combs are having a remarkably broad impact on science and technology. Integrating these systems on a photonic chip would revolutionize instrumentation, time keeping, spectroscopy, navigation and potentially create new mass-market applications. A key element of such a system-on-a-chip will be a mode-locked comb that can be self-referenced. The recent demonstration of soliton pulses from a microresonator has placed this goal within reach. However, to provide the requisite link between microwave and optical rate signals soliton generation must occur within the bandwidth of electronic devices. So far this is possible in crytalline devices, but not chip-based devices. Here, a monolithic comb that generates electronic-rate soliton pulses is demonstrated. }

The self-referenced optical frequency comb is revolutionizing a wide range of subjects spanning spectroscopy to time standards \cite{jones2000carrier,holzwarth2000optical,diddams2001optical,udem2002optical,diddams2004standards,newbury2011searching,steinmetz2008laser,li2008laser,fortier2011generation}.  Since their invention, a miniaturized approach to the formation of a comb of optical frequencies has been proposed in high-Q microresonators \cite{del2007optical,kippenberg2011microresonator}. These microcombs or Kerr combs have been demonstrated in several material systems \cite{savchenkov2008tunable,grudinin2009generation,papp2011spectral,hausmann2014diamond}, including certain planar systems suitable for monolithic integration \cite{levy2010cmos,razzari2010cmos,ferdous2011spectral,li2012low,jung2013optical}. They have been applied in demonstrations of microwave generation \cite{savchenkov2008tunable}, waveform synthesis \cite{ferdous2011spectral}, optical atomic clocks \cite{papp2014microresonator} and coherent communications \cite{pfeifle2014coherent}. Microcombs were initially realized through a process of cascaded four-wave mixing \cite{del2007optical} driven by parametric oscillation \cite{kippenberg2004kerr,savchenkov2004low}. However, a recent advance has been the demonstration of mode locking through formation of dissipative solitons \cite{herr2014temporal,2014arXiv1410.8598B}.  While initial work on microcombs demonstrated phase-locked states \cite{herr2012universal,li2012low,del2014self,del2015phase,papp2013parametric} including pulse generation\cite{saha2013modelocking}, solitons are both phase locked and, being pulses, can be readily broadened spectrally \cite{herr2014temporal}. Moreover, resonator dispersion can be engineered so as to create coherent dispersive waves that further broaden the soliton comb spectrum within the resonator \cite{2014arXiv1410.8598B,wang2014broadband}. Dissipative solitons balance dispersion with the Kerr nonlinearity while also balancing optical loss with parametric gain from the Kerr nonlinearity \cite{ankiewicz2008dissipative,herr2014temporal}. They have been observed in fiber resonator systems \cite{leo2010temporal}.  In microresonators, dissipative solitons have been observed in fluoride-based \cite{herr2014temporal, grudinin2015towards} and in silicon-nitride based microresonator systems \cite{2014arXiv1410.8598B}. Crystalline system generated solitons have also been externally broadened to 2/3 of an octave enabling detection of the comb offset frequency \cite{jost2014microwave}. In this work, soliton mode locking in silica is reported using an ultra-high-Q silica-on-silicon wedge resonator. Moreover, the soliton repetition rate is readily detectable with commercial photo-detectors. The ability to generate solitons on a chip at rates commensurate with detectors and electronics is an essential step in the ultimate goal of a fully-integrated comb system. Microfabrication also provides a high level of control of both soliton repetition rate and soliton mode family dispersion.

\begin{figure*}
  \begin{centering}
  \includegraphics[width=17.5 cm]{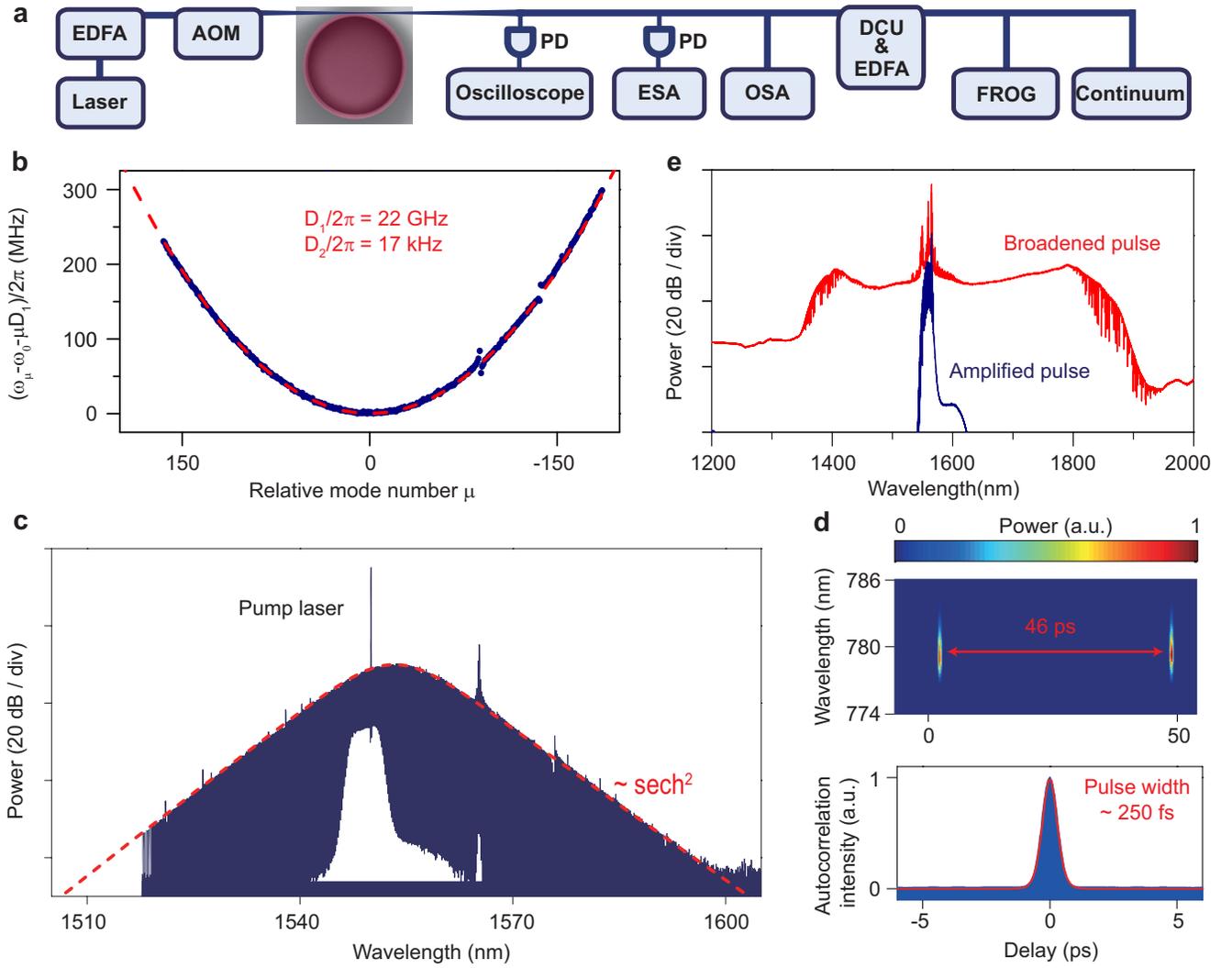}
  \captionsetup{singlelinecheck=no, justification = RaggedRight}
  \caption{\textbf{Experiment setup, soliton mode-family dispersion, optical spectrum, autocorrelation and pulse-broadening spectrum.} \textbf{a,} Experimental setup. A continuous-wave fiber laser is amplified by an erbium-doped fiber amplifier (EDFA). The laser frequency is separately monitored using a fiber Mach-Zhender interferometer (not shown). To achieve ``power kicking" the laser is modulated by an acousto-optic modulator (AOM). The laser is coupled to a high-Q wedge resonator using a fiber taper. The power transmission and RF beatnote are detected by photodetectors (PD) and sent to an oscilloscope and electrical spectrum analyzer (ESA). Soliton optical spectra are measured using an optical spectrum analyzer (OSA). A dispersion compensation unit (DCU) is employed before solitons are amplified for FROG measurement and spectral broadening. Polarization controllers, optical isolators, and optical Bragg filters are not shown. \textbf{b,} Frequency dispersion of the modes belonging to the soliton-forming mode family is plotted versus relative mode number. To construct this plot, mode frequency relative to a $\mu=0$ mode (mode to be pumped) is first measured using a calibrated Mach-Zehnder interferometer (fiber optic based). To second order in the mode number the mode frequency is given by the Taylor expansion $\omega_\mu=\omega_0+\mu \mathrm{D}_1+\frac{1}{2}\mu^2 \mathrm{D}_2$, and the red curve is a fit using parameters given in the inset.  In the plot, the mode frequencies are offset by the linear term in the Taylor expansion to make clear the second-order group dispersion. The measured modes span wavelengths from 1520 nm to 1580 nm and $\mu = 0$ corresponds to a wavelength close to 1550 nm. Non-soliton forming mode families have been removed in the data, however, their presence can be seen through perturbations to the parabolic shape. \textbf{c,} Optical spectrum of single soliton state with a $\mathrm{sech}^{2}$ envelope (red dashed line) superimposed for comparison. The pump laser is suppressed by 20 dB with an optical Bragg filter. \textbf{d,} FROG (upper) and autocorrelation trace (lower) of the soliton state in \textbf{c}. The optical pulse period is 46 ps and the fitted pulse width is 250 fs (red solid line). \textbf{e} In red: optical spectrum of the soliton pulse train after amplification using an erbium fiber amplifier and nonlinear broadening using a section of highly nonlinear fiber. Correction of dispersion in the amplifier was accomplished using a waveshaper. In blue: the unbroadened spectrum immediately after amplification has been shifted vertically for clarity. Its spectral shape is modified by the amplifier relative to the original soliton spectrum. }
  \label{fig1}
\end{centering}
\end{figure*}

Silica wedge resonators were fabricated using float-zone silicon wafers \cite{lee2012chemically}. The devices exhibit a nearly constant finesse over a wide range of diameters and have previously been applied for comb generation at free-spectral-range ($FSR$) values from 2.6 GHz to 220 GHz, including the formation of stable phase-locked comb states \cite{li2012low,papp2014microresonator}. In the present work, 3 mm diameter devices with an $FSR$ of 22 GHz were prepared and intrinsic Q factors were characterized by linewidth measurement to lie near 400 million. To both characterize the soliton tuning range (see discussion below) and to provide a separate test of the Q factor, the threshold for parametric oscillation was measured and compared to theory\cite{kippenberg2004kerr,li2012low},
\begin{equation}
\mathrm{P_{th}} = {\pi n \omega_o A_{\mathrm{eff}} \over 4 \eta n_2} {1 \over \mathrm{D}_1 Q^2 },
\end{equation}
where $A_{\mathrm{eff}}\sim 30$ $\mathrm{\mu m}^2$ is the effective mode area, $n$ is refraction index, $n_2$ is the Kerr coefficient, $\mathrm{D}_1$ is the $FSR$ in rad/s units, $\eta=Q/Q_{\mathrm{ext}}$ characterizes the waveguide to resonator loading where $Q_{\mathrm{ext}}$ is the external or coupling Q-factor and $Q$ is the total Q factor (intrinsic loss and loading included), and $\omega_o$ is the optical frequency. Measurements of threshold power slightly larger than 1 mW were consistent with the measured Q factors. 

\begin{figure*}
  \begin{centering}
  \includegraphics[width=17.5 cm]{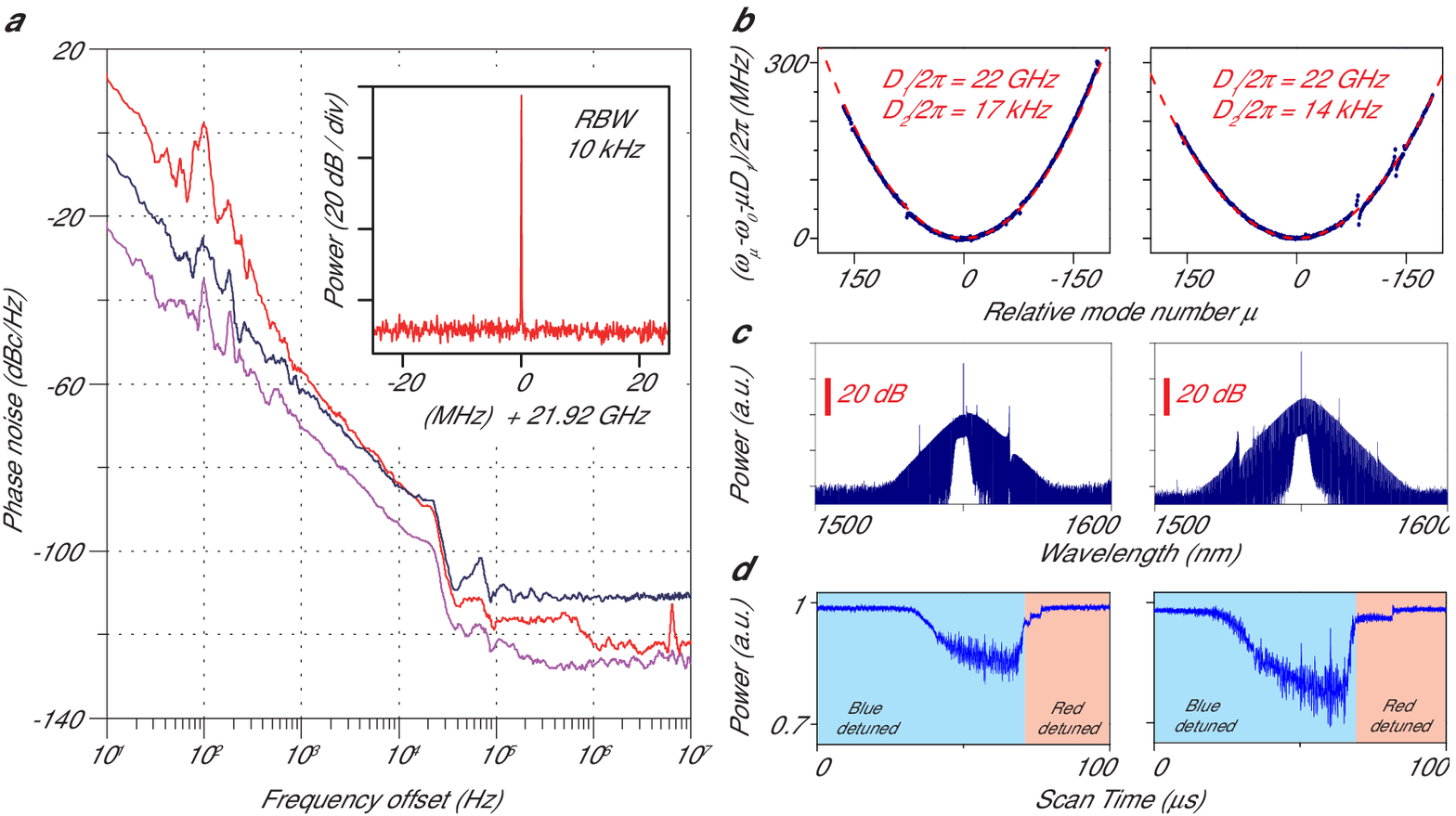}
  \captionsetup{singlelinecheck=no, justification = RaggedRight}
  \caption{\textbf{Detected phase noise and electrical spectra for three devices with corresponding mode dispersion and soliton data.} \textbf{a,}  Phase noise spectral density function versus offset frequency from the detected soliton repetition frequency of three different devices. A Rohde Schwarz phase noise analyzer was used in the measurement. Inset shows the electrical spectrum of the soliton fundamental frequency of 21.92 GHz for one device. The other devices had similar spectra with repetition frequencies of 22.01 and 21.92 GHz. \textbf{b,c,} The mode dispersion spectra and soliton spectra for two of the devices measured in \textbf{a}. The third device in \textbf{a} is from figure 1. \textbf{d,} Pump power transmission versus tuning across a resonance used to generate the soliton spectra in \textbf{c}. The data show the formation of steps as the pump tunes red relative to the resonance. Both blue-detuned and red-detuned operation of the pump relative to the resonance are inferred from generation of an error signal using a Pound-Drever-Hall system operated open loop.}
  \label{fig2}
\end{centering}
\end{figure*}

The mode family that is phase locked to form the soliton train of pulses must feature anomalous dispersion \cite{kippenberg2004kerr} and also minimal distortion of the dispersion caused by other mode families within the resonator \cite{herr2014mode}. The first of these requirements is straightforward in silica wedge resonators when operated in the 1.5 micron band \cite{li2012sideband}. However, the second requirement is more complicated as wedge resonators feature many transverse mode families. Minimizing avoided-crossing behavior is achieved by screening wedge disks to find combinations of diameter and wedge angle that produce avoided-crossing-free spectral regions. In addition, it is observed that high-Q-factor mode families are generally more immune to avoided-crossing distortion. A schematic of the experimental characterization setup is provided in figure 1a. To measure both mode family dispersion and avoided mode crossing behavior, mode frequencies were measured using a tunable laser that was calibrated with a fiber Mach-Zehnder interferometer. A measurement taken on a typical mode family used to produce solitons is presented in figure 1b.  A parabolic fit to the data featuring an anomalous dispersion of 17 kHz/$FSR$ is provided for comparison to the data. Other mode families have been removed from the spectrum for clarity, however, two avoided mode crossings are apparent in the spectrum.

Modes belonging to families featuring both anomalous dispersion and well-behaved spectra (i.e., few avoided crossings) were then pumped using a scanning fiber laser.  Solitons form when the pump frequency is red detuned relative to an optical mode and give rise to characteristic steps in the pump power transmission versus tuning \cite{herr2014temporal}. As has been described elsewhere, excitation of stable soliton trains is complicated by the thermal nonlinearity of the resonator \cite{2014arXiv1410.8598B}, which is well known to destabilize a red-detuned pump wave \cite{carmon2004dynamical}. Fortuitously, solitons feature a power dependence with detuning of the pump wave that reverses this behavior and will stabilize the pump on the red-detuning side of the resonance once the non-soliton transients have died away. To induce stability a power kicking technique similar that described in reference \cite{2014arXiv1410.8598B} was employed.  Both single and multiple soliton states were stably excited in different resonators. Figure 1c shows the spectrum measured for a single-soliton state.  The square of a hyperbolic secant function is also overlaid onto the spectrum to verify the characteristic single-soliton spectral shape. From this fitting the soliton pulse width $\tau$ is inferred to be 130 fs where the pulse shape is $\mathrm{sech}^2(t/\tau)$. The presence of small spurs in the spectrum of figure 1c correlate with the appearance of avoided crossings in the mode dispersion spectrum in figure 1b. The spectrum is slightly red-detuned relative to the pump wavelength. Displaced spectra can be produced by several mechanisms \cite{zhang2014self} including Raman self-shifting \cite{mitschke1986discovery,gordon1986theory, karpov2015raman} and soliton recoil caused by dispersive wave generation \cite{2014arXiv1410.8598B}. However, there was no evidence of dispersive wave generation at shorter wavelengths in this work, and therefore Raman self-shifting is likely the cause of this red-detuning.

Direct confirmation of single-soliton generation is provided by Frequency-Resolved Optical Gating (FROG) and autocorrelation traces (see figure 1d). In these measurements, the pump laser was suppressed by fiber Bragg filters and dispersion compensation of -1.5 ps/nm was applied using a programable optical filter before the comb was amplified by an Erbium-doped fiber amplifier (EDFA). A pulse width of 250 fs with a pulse period of 46 ps is inferred from this data. The measured pulse width is larger than that fitted from optical spectrum (130 fs) due to the limited wavelength bandwidth of the optical pre-amplifier used in this measurement. The single soliton pulses are suitable for spectral broadening. In figure 1e amplified pulses have been broadened using a nonlinear optical fiber to a spectral span of nearly 500 nm. Reducing soliton repetition rate and optimizing fiber dispersion can further increase the span. In all of these measurements, the soliton state was typically stable for several hours. 

An important feature of the soliton states generated in this work is their detectable and stable repetition rate. Figure 2a contains phase noise spectra of the detected soliton fundamental repetition frequency measured using single solitons generated with three different resonators. The upper right inset to figure 2a is a typical, radio frequency spectrum of the fundamental repetition frequency. The repetition frequency can be seen to be 21.92 GHz (resolution bandwidth is 10 kHz) and has an excellent stability that is comparable to a good K-band microwave oscillator.  For example, one of the devices measured has a phase noise level of -100 dBc/Hz at 10 kHz offset (referenced to a 10 GHz carrier frequency).  

\begin{figure}
  \includegraphics[width=8.5 cm]{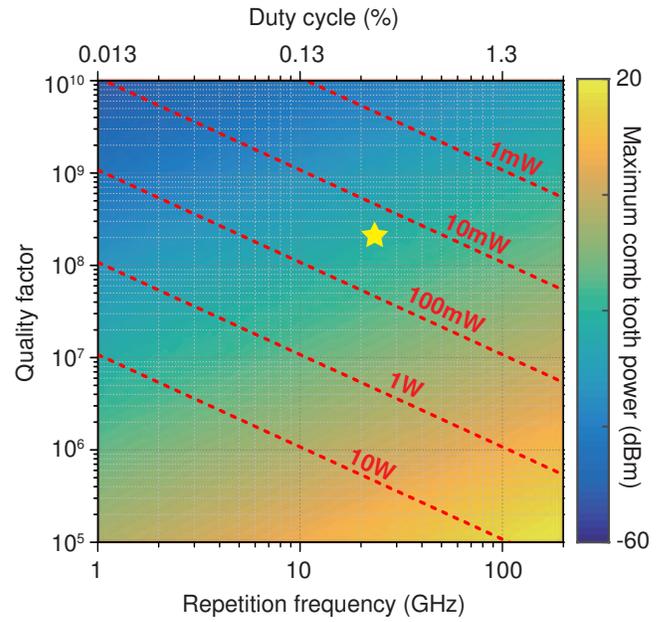}
  \captionsetup{singlelinecheck=no, justification = RaggedRight}
  \caption{\textbf{Required pumping power and maximum power per comb tooth.} Line plots are the contours of minimum pump power required for existence of 130 fs solitons in a silica resonator having the indicated Q factor and repetition frequency (note that scales change for other material systems). These are derived from the eqn. (3). Coloration indicates the maximum comb tooth power per eqn. (4). Star is estimate for this work under condition of critical coupling.}
  \label{fig3}
\end{figure}

Finally, the ability to reproduce mode family dispersion characteristics in different resonators was investigated. Figure 2b and 2c give the results of mode family dispersion measurements and soliton optical spectra on two resonators that are close in size and shape to the device in figure 1. There is high level of consistency in the both the magnitude of the dispersion and also minimal presence of mode crossings in the measured spectra. However, it is important to note that even with this consistency, there were observable variations in the nature of soliton steps formed in scanning the resonators (see figure 2d). Also, the microwave phase noise spectra in figure 2a were measured on soliton trains using these same three devices and also exhibit differences. Some of these variations could result from the other non-soliton-forming mode families that produce slightly different avoided crossing features in each of the modal spectra. 

The red-detuned location of soliton steps was confirmed using an open-loop Pound-Drever-Hall system. For steady-state soliton excitation this system was deactivated and it was not possible to directly infer the amount of pump detuning. However, the soliton pulse width $\tau$ is related to this detuning through the following approximate expression. This and subsequent expressions are valid for large normalized pump detuning\cite{herr2014temporal,wabnitz1993suppression}. 
\begin{equation}
\xi_o \equiv {2 }{\omega_o - \omega_p \over \kappa} = -{ c  \beta_2 \over n \kappa} { 1 \over \tau^2},
\end{equation}
where $\omega_p$ ($\omega_o$) is the pump (mode) frequency, $\kappa=\omega_o/Q$ is the resonance linewidth, and $\beta_2 =-{n \over c} \mathrm{D}_2 / \mathrm{D}_1^2$ is the group velocity dispersion and is negative for anomalous dispersion. Using parameters inferred from the modal dispersion measurements and the pulse times obtained from optical spectra fitting a value of $\xi_o \sim 50$ is obtained (25 MHz in absolute frequency). This value is large enough to satisfy the approximation used in the derivation of eqn. (2). Also, the pump power, $\mathrm{P_{in}}$, values typically used in this experiment were in the range of 200 mW. At these levels the predicted maximum pump detuning for soliton existence \cite{herr2014temporal,wabnitz1993suppression} is $\xi_o^{max} = \frac{\pi^2 \mathrm{P_{in}}}{8 \mathrm{P_{th}}} \sim 200$. This much wider maximum allowable detuning is consistent with the observation of very stable soliton operation. As noted, solitons were typically stable for several hours without any temperature or laser tuning stabilization. 

The generation of solitons at detectable repetition rates relies upon high resonator Q factor. Reduced soliton repetition rates are accompanied by increased mode volume, and therefore require greater pump powers to maintain nonlinear coupling of modes as is evident in eqn. (1) for parametric threshold\cite{li2012low}. The offsetting effect of Q on repetition frequency is also apparent in the expression for minimum pump power for soliton existence. This can be derived by setting $\xi_o = \xi_o^{max}$ and solving for power\cite{herr2014temporal,wabnitz1993suppression}, 
\begin{equation}
\mathrm{P}^{\mathrm{min}}_\mathrm{in}= - {2 c \over \pi} \frac{A_{\mathrm{eff}}\beta_2 }{\eta n_2  \tau^2}\frac{1}{Q\mathrm{D}_1},
\end{equation}
For a fixed pulse duration $\tau$ (equivalently fixed comb bandwidth), the required input power is inversely proportional to repetition rate and Q factor (equal to finesse $\times$ optical frequency). A plot showing linear contours of predicted pumping power versus Q factor and repetition frequency (eqn. (3)) is provided in figure 3. The plot assumes material parameters for silica and 130 fs soliton pulse width. It is important to realize that this value represents an absolute minimum. In practice a pumping level higher than this value is required for stable operation.

Besides pumping power another interesting parameter is the achievable maximum comb tooth power. The power spectral envelope for the single soliton state is given by \cite{herr2014temporal,wabnitz1993suppression},

\begin{equation}
P(\Delta\omega)= - {\pi c \over 8} \frac{ A_{\mathrm{eff}}\beta_2}{\eta n_2} \frac{\mathrm{D}_1}{Q}\mathrm{sech}^2(\frac{\pi\tau}{2} \Delta\omega),
\end{equation}
where $\Delta\omega$ denotes the comb tooth frequency relative to the pump. Note that the peak power of the spectral envelope (i.e. maximum comb tooth power) is independent of the pumping power and determined entirely by the cavity properties. The maximum tooth power varies linearly with the repetition frequency and inversely with the cavity Q factor and has been represented by the color shading in figure 3. The soliton nonlinear conversion efficiency has been described elsewhere \cite{bao2014nonlinear}.

The nearly constant finesse of the wedge resonator over a wide range of $FSR$ values \cite{lee2012chemically,li2012low} suggests that soliton repetition frequency can be widely varied at nearly constant pumping power. Accordingly, it should be possible to lower the soliton duty cycle in the present geometry by up to an order of magnitude. At a comparable average amplified power level, the peak soliton intensity would therefore increase ten-fold and the expected spectral broadening using nonlinear fiber would be much larger than observed here. Moreover, dispersion control methods have been demonstrated using a modification to the standard wedge resonator process \cite{yang2015broadband}. These same methods might be applied to control dispersive wave generation within the resonator to achieve direct generation of a broader comb. Besides soliton generation, the silica wedge resonator platform has been used to generate ultra-narrow linewidth laser sources \cite{li2012characterization,loh2015dual}, high-Q reference cavities \cite{lee2013spiral} and for continuum generation \cite{oh2014supercontinuum}. These elements are required in both self-referenced combs as well their application to clocks, high-stability microwave sources, and optical synthesizers. The results presented here therefore add to this suite of technologically compatible devices that can potentially create comb-based systems on a chip.

\vskip 0.25in

\noindent {\bf Acknowledgments} The authors thank Tobias Kippenberg, Victor Brasch at EPFL and Michael Gorodetsky at Moscow State University for helpful discussions and comments on this manuscript. The authors gratefully acknowledge the Defense Advanced Research Projects Agency under the QuASAR program and the PULSE program, the Kavli Nanoscience Institute and the Institute for Quantum Information and Matter, an NSF Physics Frontiers Center with support of the Gordon and Betty Moore Foundation.

\bibliography{soliton}

\begin{thebibliography}{10}
\expandafter\ifx\csname url\endcsname\relax
  \def\url#1{\texttt{#1}}\fi
\expandafter\ifx\csname urlprefix\endcsname\relax\def\urlprefix{URL }\fi
\providecommand{\bibinfo}[2]{#2}
\providecommand{\eprint}[2][]{\url{#2}}

\bibitem{jones2000carrier}
\bibinfo{author}{Jones, D.~J.} \emph{et~al.}
\newblock \bibinfo{title}{Carrier-envelope phase control of femtosecond
  mode-locked lasers and direct optical frequency synthesis}.
\newblock \emph{\bibinfo{journal}{Science}} \textbf{\bibinfo{volume}{288}},
  \bibinfo{pages}{635--639} (\bibinfo{year}{2000}).

\bibitem{holzwarth2000optical}
\bibinfo{author}{Holzwarth, R.} \emph{et~al.}
\newblock \bibinfo{title}{Optical frequency synthesizer for precision
  spectroscopy}.
\newblock \emph{\bibinfo{journal}{Phys. Rev. Lett.}}
  \textbf{\bibinfo{volume}{85}}, \bibinfo{pages}{2264} (\bibinfo{year}{2000}).

\bibitem{diddams2001optical}
\bibinfo{author}{Diddams, S.} \emph{et~al.}
\newblock \bibinfo{title}{An optical clock based on a single trapped
  $\mathrm{^{199} Hg^{+}}$ ion}.
\newblock \emph{\bibinfo{journal}{Science}} \textbf{\bibinfo{volume}{293}},
  \bibinfo{pages}{825--828} (\bibinfo{year}{2001}).

\bibitem{udem2002optical}
\bibinfo{author}{Udem, T.}, \bibinfo{author}{Holzwarth, R.} \&
  \bibinfo{author}{H{\"a}nsch, T.~W.}
\newblock \bibinfo{title}{Optical frequency metrology}.
\newblock \emph{\bibinfo{journal}{Nature}} \textbf{\bibinfo{volume}{416}},
  \bibinfo{pages}{233--237} (\bibinfo{year}{2002}).

\bibitem{diddams2004standards}
\bibinfo{author}{Diddams, S.}, \bibinfo{author}{Bergquist, J.},
  \bibinfo{author}{Jefferts, S.} \& \bibinfo{author}{Oates, C.}
\newblock \bibinfo{title}{Standards of time and frequency at the outset of the
  21st century}.
\newblock \emph{\bibinfo{journal}{Science}} \textbf{\bibinfo{volume}{306}},
  \bibinfo{pages}{1318--1324} (\bibinfo{year}{2004}).

\bibitem{newbury2011searching}
\bibinfo{author}{Newbury, N.~R.}
\newblock \bibinfo{title}{Searching for applications with a fine-tooth comb}.
\newblock \emph{\bibinfo{journal}{Nature Photon.}}
  \textbf{\bibinfo{volume}{5}}, \bibinfo{pages}{186--188}
  (\bibinfo{year}{2011}).

\bibitem{steinmetz2008laser}
\bibinfo{author}{Steinmetz, T.} \emph{et~al.}
\newblock \bibinfo{title}{Laser frequency combs for astronomical observations}.
\newblock \emph{\bibinfo{journal}{Science}} \textbf{\bibinfo{volume}{321}},
  \bibinfo{pages}{1335--1337} (\bibinfo{year}{2008}).

\bibitem{li2008laser}
\bibinfo{author}{Li, C.-H.} \emph{et~al.}
\newblock \bibinfo{title}{A laser frequency comb that enables radial velocity
  measurements with a precision of 1 cm s -1}.
\newblock \emph{\bibinfo{journal}{Nature}} \textbf{\bibinfo{volume}{452}},
  \bibinfo{pages}{610--612} (\bibinfo{year}{2008}).

\bibitem{fortier2011generation}
\bibinfo{author}{Fortier, T.} \emph{et~al.}
\newblock \bibinfo{title}{Generation of ultrastable microwaves via optical
  frequency division}.
\newblock \emph{\bibinfo{journal}{Nature Photon.}}
  \textbf{\bibinfo{volume}{5}}, \bibinfo{pages}{425--429}
  (\bibinfo{year}{2011}).

\bibitem{del2007optical}
\bibinfo{author}{Del'Haye, P.} \emph{et~al.}
\newblock \bibinfo{title}{Optical frequency comb generation from a monolithic
  microresonator}.
\newblock \emph{\bibinfo{journal}{Nature}} \textbf{\bibinfo{volume}{450}},
  \bibinfo{pages}{1214--1217} (\bibinfo{year}{2007}).

\bibitem{kippenberg2011microresonator}
\bibinfo{author}{Kippenberg, T.~J.}, \bibinfo{author}{Holzwarth, R.} \&
  \bibinfo{author}{Diddams, S.}
\newblock \bibinfo{title}{Microresonator-based optical frequency combs}.
\newblock \emph{\bibinfo{journal}{Science}} \textbf{\bibinfo{volume}{332}},
  \bibinfo{pages}{555--559} (\bibinfo{year}{2011}).

\bibitem{savchenkov2008tunable}
\bibinfo{author}{Savchenkov, A.~A.} \emph{et~al.}
\newblock \bibinfo{title}{Tunable optical frequency comb with a crystalline
  whispering gallery mode resonator}.
\newblock \emph{\bibinfo{journal}{Phys. Rev. Lett.}}
  \textbf{\bibinfo{volume}{101}}, \bibinfo{pages}{093902}
  (\bibinfo{year}{2008}).

\bibitem{grudinin2009generation}
\bibinfo{author}{Grudinin, I.~S.}, \bibinfo{author}{Yu, N.} \&
  \bibinfo{author}{Maleki, L.}
\newblock \bibinfo{title}{Generation of optical frequency combs with a
  $\mathrm{CaF_2}$ resonator}.
\newblock \emph{\bibinfo{journal}{Opt. Lett.}} \textbf{\bibinfo{volume}{34}},
  \bibinfo{pages}{878--880} (\bibinfo{year}{2009}).

\bibitem{papp2011spectral}
\bibinfo{author}{Papp, S.~B.} \& \bibinfo{author}{Diddams, S.~A.}
\newblock \bibinfo{title}{Spectral and temporal characterization of a
  fused-quartz-microresonator optical frequency comb}.
\newblock \emph{\bibinfo{journal}{Phys. Rev. A}} \textbf{\bibinfo{volume}{84}},
  \bibinfo{pages}{053833} (\bibinfo{year}{2011}).

\bibitem{hausmann2014diamond}
\bibinfo{author}{Hausmann, B.}, \bibinfo{author}{Bulu, I.},
  \bibinfo{author}{Venkataraman, V.}, \bibinfo{author}{Deotare, P.} \&
  \bibinfo{author}{Lon{\v{c}}ar, M.}
\newblock \bibinfo{title}{Diamond nonlinear photonics}.
\newblock \emph{\bibinfo{journal}{Nature Photon.}}
  \textbf{\bibinfo{volume}{8}}, \bibinfo{pages}{369--374}
  (\bibinfo{year}{2014}).

\bibitem{levy2010cmos}
\bibinfo{author}{Levy, J.~S.} \emph{et~al.}
\newblock \bibinfo{title}{CMOS-compatible multiple-wavelength oscillator for
  on-chip optical interconnects}.
\newblock \emph{\bibinfo{journal}{Nature Photon.}}
  \textbf{\bibinfo{volume}{4}}, \bibinfo{pages}{37--40} (\bibinfo{year}{2010}).

\bibitem{razzari2010cmos}
\bibinfo{author}{Razzari, L.} \emph{et~al.}
\newblock \bibinfo{title}{CMOS-compatible integrated optical hyper-parametric
  oscillator}.
\newblock \emph{\bibinfo{journal}{Nature Photon.}}
  \textbf{\bibinfo{volume}{4}}, \bibinfo{pages}{41--45} (\bibinfo{year}{2010}).

\bibitem{ferdous2011spectral}
\bibinfo{author}{Ferdous, F.} \emph{et~al.}
\newblock \bibinfo{title}{Spectral line-by-line pulse shaping of on-chip
  microresonator frequency combs}.
\newblock \emph{\bibinfo{journal}{Nature Photon.}}
  \textbf{\bibinfo{volume}{5}}, \bibinfo{pages}{770--776}
  (\bibinfo{year}{2011}).

\bibitem{li2012low}
\bibinfo{author}{Li, J.}, \bibinfo{author}{Lee, H.}, \bibinfo{author}{Chen, T.}
  \& \bibinfo{author}{Vahala, K.~J.}
\newblock \bibinfo{title}{Low-pump-power, low-phase-noise, and microwave to
  millimeter-wave repetition rate operation in microcombs}.
\newblock \emph{\bibinfo{journal}{Phys. Rev. Lett.}}
  \textbf{\bibinfo{volume}{109}}, \bibinfo{pages}{233901}
  (\bibinfo{year}{2012}).

\bibitem{jung2013optical}
\bibinfo{author}{Jung, H.}, \bibinfo{author}{Xiong, C.}, \bibinfo{author}{Fong,
  K.~Y.}, \bibinfo{author}{Zhang, X.} \& \bibinfo{author}{Tang, H.~X.}
\newblock \bibinfo{title}{Optical frequency comb generation from aluminum
  nitride microring resonator}.
\newblock \emph{\bibinfo{journal}{Opt. Lett.}} \textbf{\bibinfo{volume}{38}},
  \bibinfo{pages}{2810--2813} (\bibinfo{year}{2013}).

\bibitem{papp2014microresonator}
\bibinfo{author}{Papp, S.~B.} \emph{et~al.}
\newblock \bibinfo{title}{Microresonator frequency comb optical clock}.
\newblock \emph{\bibinfo{journal}{Optica}} \textbf{\bibinfo{volume}{1}},
  \bibinfo{pages}{10--14} (\bibinfo{year}{2014}).

\bibitem{pfeifle2014coherent}
\bibinfo{author}{Pfeifle, J.} \emph{et~al.}
\newblock \bibinfo{title}{Coherent terabit communications with microresonator
  Kerr frequency combs}.
\newblock \emph{\bibinfo{journal}{Nature Photon.}}
  \textbf{\bibinfo{volume}{8}}, \bibinfo{pages}{375--380}
  (\bibinfo{year}{2014}).

\bibitem{kippenberg2004kerr}
\bibinfo{author}{Kippenberg, T.}, \bibinfo{author}{Spillane, S.} \&
  \bibinfo{author}{Vahala, K.}
\newblock \bibinfo{title}{Kerr-nonlinearity optical parametric oscillation in
  an ultrahigh-Q toroid microcavity}.
\newblock \emph{\bibinfo{journal}{Phys. Rev. Lett.}}
  \textbf{\bibinfo{volume}{93}}, \bibinfo{pages}{083904}
  (\bibinfo{year}{2004}).

\bibitem{savchenkov2004low}
\bibinfo{author}{Savchenkov, A.~A.} \emph{et~al.}
\newblock \bibinfo{title}{Low threshold optical oscillations in a whispering
  gallery mode $\mathrm{CaF_2}$ resonator}.
\newblock \emph{\bibinfo{journal}{Phys. Rev. Lett.}}
  \textbf{\bibinfo{volume}{93}}, \bibinfo{pages}{243905}
  (\bibinfo{year}{2004}).

\bibitem{herr2014temporal}
\bibinfo{author}{Herr, T.} \emph{et~al.}
\newblock \bibinfo{title}{Temporal solitons in optical microresonators}.
\newblock \emph{\bibinfo{journal}{Nature Photon.}}
  \textbf{\bibinfo{volume}{8}}, \bibinfo{pages}{145--152}
  (\bibinfo{year}{2014}).

\bibitem{2014arXiv1410.8598B}
\bibinfo{author}{{Brasch}, V.} \emph{et~al.}
\newblock \bibinfo{title}{{Photonic chip based optical frequency comb using
  soliton induced Cherenkov radiation}}.
\newblock \emph{\bibinfo{journal}{ArXiv e-prints}}  (\bibinfo{year}{2014}).
\newblock \eprint{1410.8598}.

\bibitem{herr2012universal}
\bibinfo{author}{Herr, T.} \emph{et~al.}
\newblock \bibinfo{title}{Universal formation dynamics and noise of
  Kerr-frequency combs in microresonators}.
\newblock \emph{\bibinfo{journal}{Nature Photon.}}
  \textbf{\bibinfo{volume}{6}}, \bibinfo{pages}{480--487}
  (\bibinfo{year}{2012}).

\bibitem{del2014self}
\bibinfo{author}{Del'Haye, P.}, \bibinfo{author}{Beha, K.},
  \bibinfo{author}{Papp, S.~B.} \& \bibinfo{author}{Diddams, S.~A.}
\newblock \bibinfo{title}{Self-injection locking and phase-locked states in
  microresonator-based optical frequency combs}.
\newblock \emph{\bibinfo{journal}{Phys. Rev. Lett.}}
  \textbf{\bibinfo{volume}{112}}, \bibinfo{pages}{043905}
  (\bibinfo{year}{2014}).

\bibitem{del2015phase}
\bibinfo{author}{Del'Haye, P.} \emph{et~al.}
\newblock \bibinfo{title}{Phase steps and resonator detuning measurements in
  microresonator frequency combs}.
\newblock \emph{\bibinfo{journal}{Nat. Commun.}} \textbf{\bibinfo{volume}{6}}
  (\bibinfo{year}{2015}).

\bibitem{papp2013parametric}
\bibinfo{author}{Papp, S.~B.}, \bibinfo{author}{Del'Haye, P.} \&
  \bibinfo{author}{Diddams, S.~A.}
\newblock \bibinfo{title}{Parametric seeding of a microresonator optical
  frequency comb}.
\newblock \emph{\bibinfo{journal}{Opt. Express}} \textbf{\bibinfo{volume}{21}},
  \bibinfo{pages}{17615--17624} (\bibinfo{year}{2013}).

\bibitem{saha2013modelocking}
\bibinfo{author}{Saha, K.} \emph{et~al.}
\newblock \bibinfo{title}{Modelocking and femtosecond pulse generation in
  chip-based frequency combs}.
\newblock \emph{\bibinfo{journal}{Opt. Express}} \textbf{\bibinfo{volume}{21}},
  \bibinfo{pages}{1335--1343} (\bibinfo{year}{2013}).

\bibitem{wang2014broadband}
\bibinfo{author}{Wang, S.}, \bibinfo{author}{Guo, H.}, \bibinfo{author}{Bai,
  X.} \& \bibinfo{author}{Zeng, X.}
\newblock \bibinfo{title}{Broadband Kerr frequency combs and intracavity
  soliton dynamics influenced by high-order cavity dispersion}.
\newblock \emph{\bibinfo{journal}{Opt. Lett.}} \textbf{\bibinfo{volume}{39}},
  \bibinfo{pages}{2880--2883} (\bibinfo{year}{2014}).

\bibitem{ankiewicz2008dissipative}
\bibinfo{author}{Ankiewicz, A.} \& \bibinfo{author}{Akhmediev, N.}
\newblock \emph{\bibinfo{title}{Dissipative Solitons: From Optics to Biology
  and Medicine}} (\bibinfo{publisher}{Springer}, \bibinfo{year}{2008}).

\bibitem{leo2010temporal}
\bibinfo{author}{Leo, F.} \emph{et~al.}
\newblock \bibinfo{title}{Temporal cavity solitons in one-dimensional Kerr
  media as bits in an all-optical buffer}.
\newblock \emph{\bibinfo{journal}{Nature Photon.}}
  \textbf{\bibinfo{volume}{4}}, \bibinfo{pages}{471--476}
  (\bibinfo{year}{2010}).

\bibitem{grudinin2015towards}
\bibinfo{author}{Grudinin, I.~S.} \& \bibinfo{author}{Yu, N.}
\newblock \bibinfo{title}{Towards efficient octave-spanning comb with
  micro-structured crystalline resonator}.
\newblock In \emph{\bibinfo{booktitle}{SPIE LASE}},
  \bibinfo{pages}{93430F--93430F} (\bibinfo{organization}{International Society
  for Optics and Photonics}, \bibinfo{year}{2015}).

\bibitem{jost2014microwave}
\bibinfo{author}{Jost, J.} \emph{et~al.}
\newblock \bibinfo{title}{Microwave to optical link using an optical
  microresonator}.
\newblock \emph{\bibinfo{journal}{arXiv preprint arXiv:1411.1354}}
  (\bibinfo{year}{2014}).

\bibitem{lee2012chemically}
\bibinfo{author}{Lee, H.} \emph{et~al.}
\newblock \bibinfo{title}{Chemically etched ultrahigh-Q wedge-resonator on a
  silicon chip}.
\newblock \emph{\bibinfo{journal}{Nature Photon.}}
  \textbf{\bibinfo{volume}{6}}, \bibinfo{pages}{369--373}
  (\bibinfo{year}{2012}).

\bibitem{herr2014mode}
\bibinfo{author}{Herr, T.} \emph{et~al.}
\newblock \bibinfo{title}{Mode spectrum and temporal soliton formation in
  optical microresonators}.
\newblock \emph{\bibinfo{journal}{Phys. Rev. Lett.}}
  \textbf{\bibinfo{volume}{113}}, \bibinfo{pages}{123901}
  (\bibinfo{year}{2014}).

\bibitem{li2012sideband}
\bibinfo{author}{Li, J.}, \bibinfo{author}{Lee, H.}, \bibinfo{author}{Yang,
  K.~Y.} \& \bibinfo{author}{Vahala, K.~J.}
\newblock \bibinfo{title}{Sideband spectroscopy and dispersion measurement in
  microcavities}.
\newblock \emph{\bibinfo{journal}{Opt. Express}} \textbf{\bibinfo{volume}{20}},
  \bibinfo{pages}{26337--26344} (\bibinfo{year}{2012}).

\bibitem{carmon2004dynamical}
\bibinfo{author}{Carmon, T.}, \bibinfo{author}{Yang, L.} \&
  \bibinfo{author}{Vahala, K.}
\newblock \bibinfo{title}{Dynamical thermal behavior and thermal self-stability
  of microcavities}.
\newblock \emph{\bibinfo{journal}{Opt. Express}} \textbf{\bibinfo{volume}{12}},
  \bibinfo{pages}{4742--4750} (\bibinfo{year}{2004}).

\bibitem{zhang2014self}
\bibinfo{author}{Zhang, L.}, \bibinfo{author}{Lin, Q.},
  \bibinfo{author}{Kimerling, L.~C.} \& \bibinfo{author}{Michel, J.}
\newblock \bibinfo{title}{Self-frequency shift of cavity soliton in Kerr
  frequency comb}.
\newblock \emph{\bibinfo{journal}{arXiv preprint arXiv:1404.1137}}
  (\bibinfo{year}{2014}).

\bibitem{mitschke1986discovery}
\bibinfo{author}{Mitschke, F.~M.} \& \bibinfo{author}{Mollenauer, L.~F.}
\newblock \bibinfo{title}{Discovery of the soliton self-frequency shift}.
\newblock \emph{\bibinfo{journal}{Opt. Lett.}} \textbf{\bibinfo{volume}{11}},
  \bibinfo{pages}{659--661} (\bibinfo{year}{1986}).

\bibitem{gordon1986theory}
\bibinfo{author}{Gordon, J.~P.}
\newblock \bibinfo{title}{Theory of the soliton self-frequency shift}.
\newblock \emph{\bibinfo{journal}{Opt. Lett.}} \textbf{\bibinfo{volume}{11}},
  \bibinfo{pages}{662--664} (\bibinfo{year}{1986}).

\bibitem{karpov2015raman}
\bibinfo{author}{Karpov, M.} \emph{et~al.}
\newblock \bibinfo{title}{Raman induced soliton self-frequency shift in
  microresonator Kerr frequency combs}.
\newblock \emph{\bibinfo{journal}{arXiv preprint arXiv:1506.08767}}
  (\bibinfo{year}{2015}).

\bibitem{wabnitz1993suppression}
\bibinfo{author}{Wabnitz, S.}
\newblock \bibinfo{title}{Suppression of interactions in a phase-locked soliton
  optical memory}.
\newblock \emph{\bibinfo{journal}{Opt. Lett.}} \textbf{\bibinfo{volume}{18}},
  \bibinfo{pages}{601--603} (\bibinfo{year}{1993}).

\bibitem{bao2014nonlinear}
\bibinfo{author}{Bao, C.} \emph{et~al.}
\newblock \bibinfo{title}{Nonlinear conversion efficiency in Kerr frequency
  comb generation}.
\newblock \emph{\bibinfo{journal}{Optics letters}}
  \textbf{\bibinfo{volume}{39}}, \bibinfo{pages}{6126--6129}
  (\bibinfo{year}{2014}).

\bibitem{yang2015broadband}
\bibinfo{author}{Yang, K.~Y.} \emph{et~al.}
\newblock \bibinfo{title}{Broadband dispersion engineered microresonator
  on-a-chip}.
\newblock \emph{\bibinfo{journal}{arXiv preprint arXiv:1506.07157}}
  (\bibinfo{year}{2015}).

\bibitem{li2012characterization}
\bibinfo{author}{Li, J.}, \bibinfo{author}{Lee, H.}, \bibinfo{author}{Chen, T.}
  \& \bibinfo{author}{Vahala, K.~J.}
\newblock \bibinfo{title}{Characterization of a high coherence, Brillouin
  microcavity laser on silicon}.
\newblock \emph{\bibinfo{journal}{Opt. Express}} \textbf{\bibinfo{volume}{20}},
  \bibinfo{pages}{20170--20180} (\bibinfo{year}{2012}).

\bibitem{loh2015dual}
\bibinfo{author}{Loh, W.} \emph{et~al.}
\newblock \bibinfo{title}{Dual-microcavity narrow-linewidth Brillouin laser}.
\newblock \emph{\bibinfo{journal}{Optica}} \textbf{\bibinfo{volume}{2}},
  \bibinfo{pages}{225--232} (\bibinfo{year}{2015}).

\bibitem{lee2013spiral}
\bibinfo{author}{Lee, H.} \emph{et~al.}
\newblock \bibinfo{title}{Spiral resonators for on-chip laser frequency
  stabilization}.
\newblock \emph{\bibinfo{journal}{Nat. Commun.}} \textbf{\bibinfo{volume}{4}}
  (\bibinfo{year}{2013}).

\bibitem{oh2014supercontinuum}
\bibinfo{author}{Oh, D.~Y.} \emph{et~al.}
\newblock \bibinfo{title}{Supercontinuum generation in an on-chip silica
  waveguide}.
\newblock \emph{\bibinfo{journal}{Opt. Lett.}} \textbf{\bibinfo{volume}{39}},
  \bibinfo{pages}{1046--1048} (\bibinfo{year}{2014}).

\end{thebibliography}

\end{document}